\documentclass[twoside]{JHEP3mod}

\usepackage{graphicx}
\usepackage{amsmath}

\title{Diquark condensation at strong coupling}

\author{V.~Azcoiti \\ 
Departamento de F\'{\i}sica Te\'orica, Universidad de
Zaragoza, Cl. Pedro Cerbuna 12, E-50009 Zaragoza (Spain) \\
E-mail: \email{azcoiti@azcoiti.unizar.es}} 
\author{G.~Di~Carlo \\ 
INFN, Laboratori Nazionali del Gran Sasso, 
67010 Assergi,(L'Aquila) (Italy) \\
E-mail: \email{dicarlo@lngs.infn.it}} 
\author{A.~Galante \\
INFN, Laboratori Nazionali del Gran Sasso, 
67010 Assergi,(L'Aquila) (Italy) and
Dipartimento di Fisica dell'Universit\`a di L'Aquila,
67100 L'Aquila (Italy) \\
E-mail: \email{galante@lngs.infn.it}}
\author{V.~Laliena \\
Departamento de F\'{\i}sica Te\'orica, Universidad de
Zaragoza, Cl. Pedro Cerbuna 12, E-50009 Zaragoza (Spain) \\
E-mail: \email{laliena@unizar.es}}

\abstract{
The possibility of diquark condensation at sufficiently large baryon
chemical potential and zero temperature is analyzed in QCD at strong 
coupling. In agreement with other strong coupling analysis,
it is found that a first order phase transition separates
a low density phase with chiral symmetry spontaneously
broken from a high density phase where chiral symmetry is restored.
In none of the phases diquark condensation takes place as an equilibrium
state, but, for any value of the chemical potential, there is a metastable 
state characterized by a non-vanishing diquark condensate. 
The energy difference between this metastable state and the equilibrium
state decreases with the chemical potential and is minimum in the high
density phase. The results indicate that there is attraction in the 
quark-quark sector also at strong coupling, and that the attraction is
more effective at high baryon density, but for infinite coupling it is
not enough to produce diquark condensation. It is argued that the 
absence of diquark condensation is not a peculiarity of the strong coupling
limit, but persists at sufficiently large finite couplings. 
}


\begin{document}


\section{Introduction}

Hadron phenomenology suggests that quarks confined within the baryons
are strongly clustered into diquarks \cite{phen1,phen2}. 
Theoretically, a weak coupling analysis of QCD at the level of one 
gluon exchange 
shows that the quark-quark scattering amplitude is attractive in the
color anti-triplet channel and repulsive in the symmetric color sextet 
channel. The attraction in the color anti-triplet channel due to
one gluon exchange provides a mechanism for bounding the diquark in
a state the color of which is neutralized by the remaining quark within
the baryon. The quark-quark interaction induced by non-perturbative
instanton effects produces also attraction in the color anti-triplet
channel and repulsion in the color sextet channel.

The effect of the quark-quark attraction should be more dramatic
at high baryon densities \cite{bailin}. At zero density 
the loosely bound diquarks are in its turn 
bound to quarks producing the color singlet baryons.
At high baryon density, however, the color charge is expected to
be deconfined and the weakly interacting quarks may form a Fermi
surface that the quark-quark attraction will render unstable. Then,
the phenomenon of color superconductivity would take place via
a BCS mechanism, with the formation of a diquark 
condensate \cite{bailin,shuryak,alford}. The diquark condensate
can be studied in QCD in the weak coupling regime via the
Schwinger-Dyson equations \cite{son,schaf:wil,pisarsky,cristina} 
and instabilities of vertex functions \cite{vertex}. 
The diquark condensation
offers the possibility of exotic phenomena at high baryon density,
as color-flavor locking and unlocking and crystalline superconductivity
\cite{alford:rew,al:raj:wil,al:ber:raj,al:bow:raj}, and the 
continuity of the quark and hadronic phases \cite{continuity}.

The diquark operator carries color charge and is not gauge invariant.
Hence, general arguments forbide diquark condensation in the naive
way \cite{elitzur} (see also \cite{splittorff}). 
Diquark condensation must be understood not as the spontaneous
breaking of the gauge symmetry, but as a kind of Higgs mechanism that
has the observable consequencies of color superconductivity 
\cite{wilczek,langfeld}.

Recently, strong evidence of diquark condensation in two colors QCD
has been found 
\cite{aloisio1,ko:to:sin,morrison,zhit,sp:son:st,sp:to:ver}. 
The analysis in this case is simpler
than in three colors QCD for two reasons: there is a gauge invariant 
diquark operator and the fermionic determinant is positive at finite 
chemical potential, so that Monte Calo simulations are feasible.

In three colors QCD, however, the arguments in favour of diquark 
condensation at high density are based on weak
coupling analysis (although nonperturbative effects are taken
into account via instantons), that should be valid at asymptotically
large densities. It is very important to verify
the diquark condensation relaxing the weak coupling assumptions. 
Unfortunately, numerical
simulations of lattice QCD at finite density are unfeasible
ought to the sign problem. Hence, the strong coupling techniques
are valuable to get insight in the problem.  Recent attempts to study
QCD at finite density in the strong coupling limit by using
Hamiltonian techniques have been developed in \cite{svet,umino,xiang}.
In this paper, we shall 
study the possibility of diquark condensation in the strong coupling
limit using the path integral formalism.

\section{Effective action at strong coupling}

Let us consider QCD regularized on a euclidean four dimensional
lattice with one flavor of staggered fermions.
The gauge group is SU(3). Let us define the
meson and baryon operators:
\begin{eqnarray}
M(x) &=& \bar\psi^a(x)\psi^a(x) \\
B(x) &=&\frac{1}{6}\epsilon_{abc}\psi^a(x)\psi^b(x)\psi^c(x) \\
\bar{B}(x) &=&\frac{1}{6}\epsilon_{abc}\bar\psi^a(x)\bar\psi^b(x)\bar\psi^c(x) 
\end{eqnarray}
where summation over repeated color indexes, $a,b,c$, is understood. 
At strong coupling,
$1/g^2=0$, the gauge field can be integrated out and gives the 
following effective action for the fermions \cite{morel}:
\begin{eqnarray}
S^\mathrm{eff}_\mathrm{sc} &=& m_0\sum_x M(x)-
\frac{1}{2}\sum_{x,y}M(x)V_\textrm{M}(x,y)M(y)-
\sum_{x,y}\bar{B}(x)V_\textrm{B}(x,y)B(y) \nonumber \\
&+& \frac{1}{576}\sum_{x,\nu}M^2(x) M^2(x+\nu)-
\frac{5}{576}\sum_{x,\nu}\bar{B}(x)B(x)\bar{B}(x+\nu)B(x+\nu) \, ,
\label{scaction}
\end{eqnarray}
where $m_0$ is the bare fermion mass and
\begin{eqnarray}
V_\textrm{M}(x,y) &=&
\frac{1}{12}\sum_\nu(\delta_{y,x+\nu}\,+\,\delta_{y,x-\nu})\, , \\
V_\textrm{B}(x,y)&=&
\frac{1}{8}\sum_\nu[f_\nu(x)\delta_{y,x+\nu}\,-
\,f_\nu(x)^{-1}\delta_{y,x-\nu}]\, , 
\end{eqnarray}
where $\nu$ are the unit vectors in the four space time directions and
\begin{equation}
f_\nu(x)\;=\left\{
\begin{array}{ll}
e^{3\mu}, & \nu=0 \\
\eta_\nu(x), & \nu=1,2,3
\end{array}
\right.
\end{equation}
where $\mu$ is the baryon chemical potential and $\eta_\nu(x)$ are the 
Kogut-Susskind phases.

The zero temperature grand canonical partition function is given by
\begin{equation}
\mathcal{Z}=\int[d\bar\psi d\psi]\,\exp[-S_\mathrm{sc}]\, .
\end{equation}

We will neglect the last two terms of the fermion effective action,
which involve eight-fermion and twelve-fermion vertices. It has been
proved that these terms are sub-dominant in a $1/d$ expansion, where
$d$ is the space-time dimension \cite{morel}. In $d=4$ their
contribution to the thermodynamics is small and does not change 
qualitatively the phase diagram \cite{dagotto}. We linearize the remaining 
meson and baryon terms in the usual way, introducing a bosonic field,
$\varphi(x)$, and two fermion fields, $b_x$ and $\bar{b}_x$, via
the Hubbard-Stratonovich transformations:
\begin{equation}
\exp\{\frac{1}{2}(M,V_\textrm{M}M)\} = (\det V_M)^{-1/2} 
\int[d\varphi]
\exp\{-\frac{1}{2}(\varphi,V^{-1}_\textrm{M}\varphi)
-(\varphi,M)\} \, , 
\end{equation}
\begin{equation}
\exp\{(\bar{B},V_\textrm{B}B)\} = \det V_B \int[d\bar{b}db]
\exp\{-\frac{1}{2}(\bar{b}, V^{-1}_\textrm{B}b)\,+\,
(\bar{b},B)+(\bar{B},b)\}\, .
\label{linbaryon}
\end{equation} 
where we use the notation $(f,A g)=\sum_{xy}f(x) A(x,y)g(y)$ and 
$(f,g)=\sum_x f(x) g(x)$.

Let us introduce the diquark fields
\begin{eqnarray}
D^a(x) &=& \epsilon_{abc}\psi^b(x)\psi^c(x) \label{diquark} \\
\bar{D}^a(x) &=& \epsilon_{abc}\bar\psi^b(x)\bar\psi^c(x) 
\label{adiquark} \, . 
\end{eqnarray}
Note that $\bar{D}^a(x)$ transforms as a quark (color triplet)
and $D^a(x)$ as an antiquark (color anti-triplet)
under gauge transformations. It is convenient to our purpose to
rewrite the term $\exp[(\bar{b},B)+(\bar{B},b)]$ entering
equation~(\ref{linbaryon})
by using the following identity:
\begin{eqnarray}
\exp[\bar{b}_xB(x)+\bar{B}(x)b_x] &=&
\exp[\frac{1}{36}\bar{b}_x b_x M(x)+2M^2(x)]\,
\int d\phi^\dagger(x) d\phi(x) \exp\left\{-\phi^\dagger_a(x)\phi_a(x)\right.
\nonumber \\
&& \left.
-\phi_a^\dagger(x)[\frac{1}{6}\bar{b}_x\psi^a(x)+\bar{D}^a(x)] 
-\phi_a(x)[\frac{1}{6}\bar\psi^a(x)b_x+D^a(x)]\right\},
\end{eqnarray}
where $\phi_a(x)$ is a complex bosonic field that carries color charge in the
fundamental representation.
The above series of transformations lead to the following expression
for the grand canonical partition function:
\begin{equation}
\mathcal{Z}=\int[d\bar\psi d\psi][d\bar{b}db][d\varphi]
[d\phi^\dagger d\phi]\,\exp[-S^\mathrm{eff}_\mathrm{fb}]\, ,
\end{equation}
where
\begin{eqnarray}
S^\mathrm{eff}_\mathrm{fb} &=& \sum_x\phi_a^\dagger(x)\phi_a(x) + 
\frac{1}{2}\sum_{x,y}\varphi(x)V^{-1}_\textrm{M}(x,y)\varphi(y) +
\sum_{x,y}\bar{b}_x V^{-1}_\textrm{B}(x,y) b_y -\ln\det V_B \nonumber \\
&+& \sum_x\left\{(m+\varphi(x)-\frac{1}{36}\bar{b}_xb_x) M(x)-2 M^2(x)
\right. \nonumber \\
&&\left. +\phi^\dagger_a(x)[\frac{1}{6}\bar{b}_x\psi^a(x)+\bar{D}^a(x)]
+\phi_a(x)[\frac{1}{6} \bar\psi^a(x) b_x+D^a(x)]\right\}\, .
\end{eqnarray}
Notice that the scalar field $\phi_a(x)$ acts as a Higgs field with
color charge in the fundamental representation.

The integral over the fermion fields $\psi^a(x)$ and $\bar\psi^a(x)$
factorizes as a product of integrals at each lattice site and can be
readily performed. Afterwards, the auxiliary fields $b_x$ and $\bar{b}_x$
can be integrated out, and we arrive to a representation of the grand
canonical partition function as a functional integral over bosonic
fields:
\begin{equation}
\mathcal{Z}=\int [d\varphi][d\phi^\dagger d\phi]\,
\exp[-S^\mathrm{eff}_\mathrm{b}]\, ,
\end{equation}
with
\begin{equation}
S^\mathrm{eff}_\mathrm{b} = \sum_x\phi_a^\dagger(x)\phi_a(x) + 
\frac{1}{2}\sum_{x,y}\varphi(x)V^{-1}_\textrm{M}(x,y)\varphi(y)
-\ln\det[\tilde\Theta_1-V^{-1}_\mathrm{B}\tilde\Theta_2]\, ,
\label{seffb}
\end{equation}
where $\tilde\Theta_i(x,y)=\Theta_i(x)\delta_{xy}$,
$i=1,2$, with
\begin{eqnarray}
\Theta_1(x) &=& m(x)[m^2(x)-4|\phi(x)|^2+12] \, , \\
\Theta_2(x) &=& \frac{1}{36}[4|\phi(x)|^4+m^2(x)|\phi(x)|^2+8|\phi(x)|^2
-3m^2(x)-12] \, .
\end{eqnarray}
In the above equation $m(x)=m_0+\varphi(x)$.

The field $\varphi(x)$ is related to the chiral condensate
in the usual way, by
\begin{equation}
\langle M(x) \rangle = c \langle \varphi(x) \rangle \, ,
\end{equation}
where $c=(1/V) \sum_{xy} V^{-1}_\textrm{M}(x,y)$, with $V$ the
lattice volume. Hence, a non vanishing vacuum expectation value
of $\varphi(x)$ implies spontaneous chiral symmetry breaking.

It is not completely obvious how the field $\phi_a(x)$ is related to the
diquark operator. One way to see it is to couple sources to the diquark
field and relate the derivatives of the partition function with
respect to the diquark sources to $\phi_a(x)$.

\section{Sources for the diquark field}

The diquark fields defined by equations~(\ref{diquark}) 
and~(\ref{adiquark}) are not gauge invariant. 
Adding sources for these fields will make sense
once a gauge is fixed. To get the effective action in the presence 
of sources, however, we do not need to fix the gauge. It can be fixed
afterwords, when using the effective action to make computations.
Let us add to the action~(\ref{scaction}) a source term of the
form
\begin{equation}
\sum_x[J_a(x) D^a(x) + J^\dagger_a(x) \bar{D}^a(x)] \, .
\end{equation}
With the same algebraic manipulations of the previous section we
find that the only effect of the sources is to replace the functions
$\Theta_1(x)$ and $\Theta_2(x)$ entering equation~(\ref{seffb}) by
\begin{eqnarray}
\Theta^{(J)}_1(x) &=& \Theta_1(x)
+4m(x)[\phi^\dagger_a(x)J_a(x)+J^\dagger_a(x)\phi_a(x)-J^\dagger_a(x)J_a(x)] 
\, , \\
\Theta^{(J)}_2(x) &=& \Theta_2(x)
-\frac{1}{9}\left\{\phi^\dagger_a(x)J_a(x)+J^\dagger_a(x)\phi_a(x)-
J^\dagger_a(x)J_a(x)\right. \nonumber \\
&& \left. +|\phi(x)|^2[\phi^\dagger_a(x)J_a(x)+J^\dagger_a(x)\phi_a(x)]
-\phi^\dagger_a(x) J_a(x)\,J^\dagger_b(x)\phi_b(x)\right\} 
 \, .
\end{eqnarray}

Taking the derivative respect to $J_a(x)$ and letting the sources vanish,
we get the diquark
operator expressed in terms of the bosonic fields $\varphi$ and
$\phi_a$:
\begin{equation}
D^a(x) = [4 \varphi(x) A(x) - \frac{1}{9}(1-|\phi(x)|^2)B(x)]\,
\phi^\dagger_a(x)\, ,
\end{equation}
where
\begin{eqnarray}
A(x) &=& \left(\frac{1}{\tilde\Theta_1-V_B\tilde\Theta_2}\right)_{xx} \, , \\
B(x) &=& \left(\frac{1}{\tilde\Theta_1-V_B\tilde\Theta_2}V_B\right)_{xx} \, .
\end{eqnarray}
In the above expressions matrix multiplication and inversion is understood.

We see, as expected, that the diquark field is proportional to
$\phi^\dagger_a$.

\section{Phase diagram}

The effective potential, $\mathcal{U}_\mathrm{eff}$, to the lowest 
order semi-classical expansion
(mean field approximation) is given by the effective action
(\ref{seffb}) evaluated at constant fields, $\varphi(x)=\bar\varphi$ and 
$|\phi(x)|^2=v^2$. In this case the determinant entering 
equation~(\ref{seffb}) can be readily computed by using the Fourier
transformation and we have
\begin{equation}
\mathcal{U}_\mathrm{eff} = v^2+\frac{3}{4}\bar\varphi^2
-\frac{1}{2}\int\frac{d^4k}{(2\pi)^4}
\ln[\bar\Theta_1^2-\bar\Theta_2^2 R^2(k)]\, ,
\label{pot}
\end{equation}
where
\begin{equation}
R^2(k)\;=\;\frac{1}{32}\left[-\sum_{\nu=1}^3 \sin^2(k_\nu/2)-2+
e^{6\mu} e^{\mathrm{i} k_0}+e^{-6\mu} e^{-\mathrm{i} k_0}
\right]\, ,
\end{equation}
and
\begin{eqnarray}
\bar\Theta_1 &=& \bar{m}(\bar{m}^2-4v^2+12) \\
\bar\Theta_2 &=& \frac{1}{36}[4v^4+v^2\bar{m}^2+8v^2-3\bar{m}^2-12]\, ,
\end{eqnarray} 
with $\bar{m}=m_0+\bar\varphi$.
A non vanishing expectation value, $v$, of the Higgs field will signal,
as usual, the presence of the Higgs phenomenon, which, in this context, 
implies the diquark condensation.

For the remaining of the paper we will restrict our analysis to
the chiral limit, $m_0=0$.

Let us analyze the phase structure as a function of the chemical potential.
The effective potential (\ref{pot}) is invariant under the change
$\bar\varphi\rightarrow -\bar\varphi$ (and notice that, by definition,
$v\geq0$).
The effective potential (\ref{pot}) has the following local minima, 
\begin{equation}
\begin{array}{lll}
v=0\, , & \bar\varphi=\pm\left(\sqrt{33}-5\right)^{1/2} 
\hspace{0.5truecm} & \forall \; \mu \\
v=\sqrt[4]{5}\, , \hspace{0.5truecm} & \bar\varphi=0 & \forall \; \mu \\
v=0\, , & \bar\varphi=0 & \mathrm{for} \;\; \mu > 0.4416 
\end{array}
\end{equation}
which are independent of $\mu$.

There is a first order phase transition at $\mu_c\approx 1.557$. For
$\mu<\mu_c$ the (degenerate ought to chiral symmetry) absolute minima are 
at $v=0$ and $\bar\varphi=\pm(\sqrt{33}-5)^{1/2}$, corresponding to a phase
with chiral symmetry spontaneously broken. For $\mu>\mu_c$ the absolute
minimum is at $v=0$ and $\bar\varphi=0$; chiral symmetry is restored in 
this phase. Obviously, the transition is first order. For $\mu<\mu_c$
the baryon density is zero, and three for $\mu>\mu_c$. Three quarks
per lattice site is the maximum allowed by Pauli's principle, so that
the transition separates a phase of zero baryon density from a phase
saturated of quarks. The system at any intermediate density is 
thermodynamically unstable and splits into domains of zero and
saturated baryon density.
This behavior is an artifact of the strong coupling limit and
has been observed in other strong coupling analysis at zero
temperature and finite chemical potential
\cite{xiang,bilic,chavel,aloisio2}.

The minimum at $v=\sqrt[4]{5}$ and $\bar\varphi=0$ is metastable in 
both phases. It describes a chiral symmetric state with a diquark 
condensate. The baryon density that correspond to this metastable
state is a smooth function of the chiral condensate 
(see figure~\ref{fig:density})
that interpolates between zero density and saturation.
This means that at any baryon density a state with diquark condensation
can be formed and have some short life until the system splits into
its zero density and saturated domains. The presence
of the metastable state signals the attraction in the quark-quark channel.
At strong coupling such attraction is not strong enough to form a stable 
diquark condensate. The energy difference between this metastable
state and the equilibrium state decreases with $\mu$,
indicating that the quark-quark attraction becomes the more effective the
higher the baryon density (see figure~\ref{fig:barrier}). However,
the energy difference is nonzero even at high $\mu$ (in the saturated 
phase).

Had we computed the minimum of the effective potential with diquark sources,
we would have obtained as metastable minimum in the limit of vanishing 
sources 
\begin{equation}
\phi_a=v\,\frac{J_a}{|J|} \, .
\end{equation}
Notice that for vanishing sources $J_a/|J|$ merely selects a direction
in color space.
This does not mean that gauge symmetry is spontaneously broken, since
the fluctuations of the redundant gauge variables will destroy the
behavior of the classical level. However, if we fix a gauge before
including the fluctuations, then the diquark field will have a
non vanishing expectation value and the Higgs mechanism will take
place. 

\FIGURE{
\centerline{\includegraphics*[width=3in,angle=90]{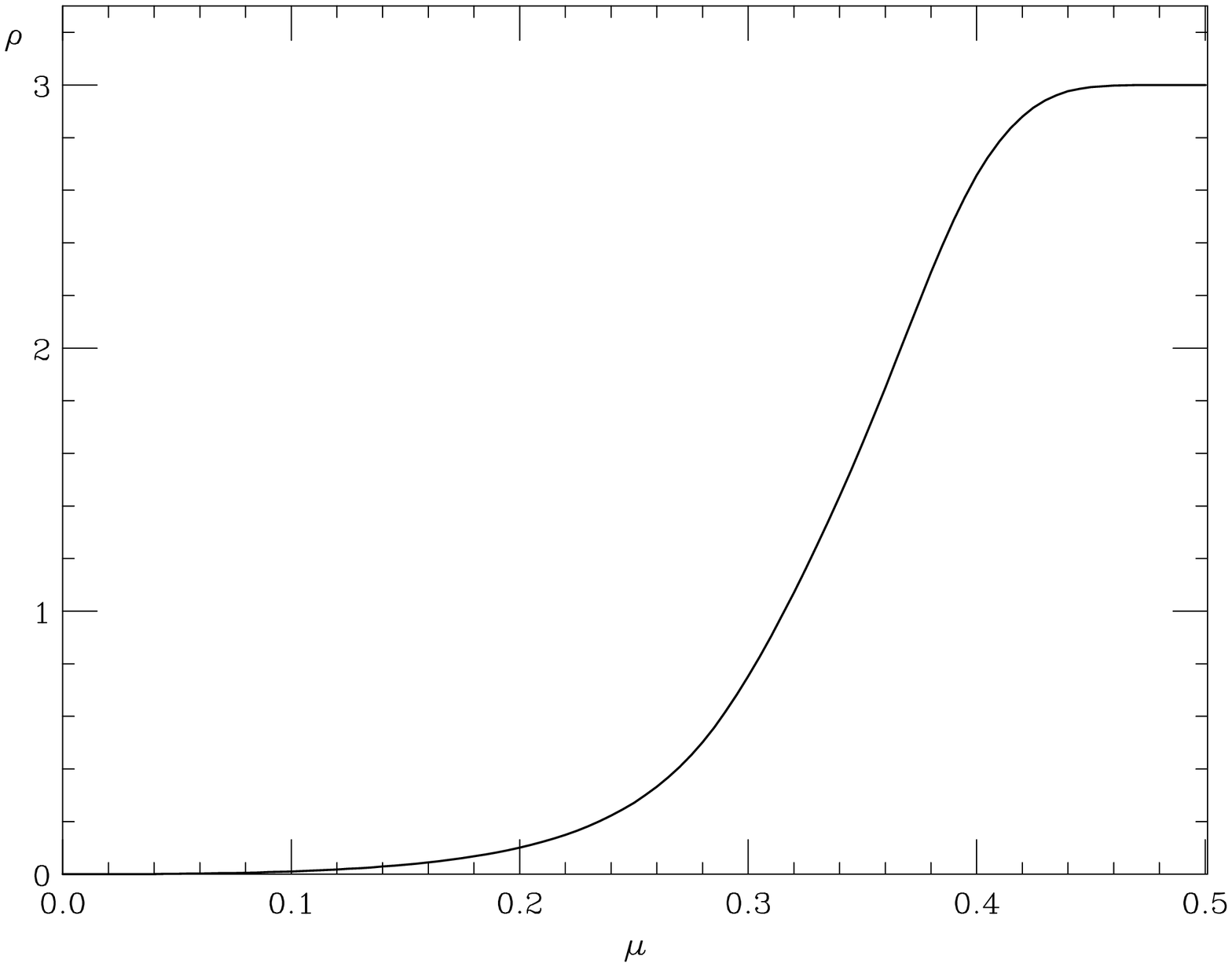}}
\caption{Baryon density as a function of the chemical potential in
the metastable state with diquark condensation.}
\label{fig:density}
}

\FIGURE{
\centerline{\includegraphics*[width=3in,angle=90]{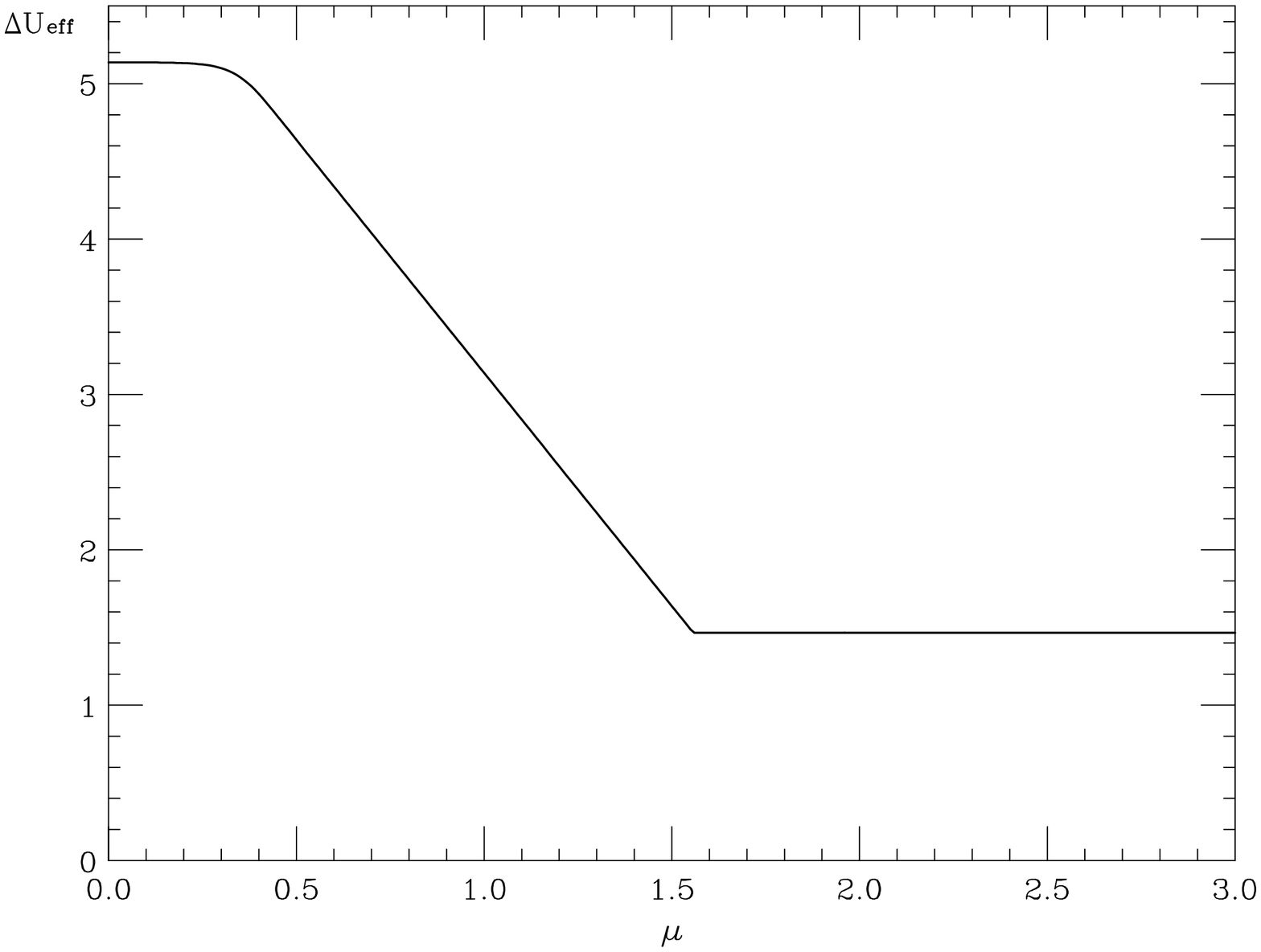}}
\caption{Energy difference between the diquark condensate metastable 
and the equilibrium state.}
\label{fig:barrier}
}

\section{Discussion}
 
It is rather natural that diquarks do not condense at strong coupling, 
since the confining forces are so strong that produce almost
pointlike baryons and mesons. The transition driven by the chemical 
potential separates
a phase with zero baryon density from a saturated phase, with three quarks
per lattice site. Hence, the system at any finite density is unstable
and splits into domains of zero and saturated densities. Deconfinement
does not take place at any density and diquark condensation cannot take 
place in a confined phase. 

It is however interesting that a metastable state in which diquark 
condensation takes place appears at any density. This shows that there
is a strong attraction in the quark-quark channel at strong coupling too,
where the weak coupling arguments based on one gluon exchange or
instanton induced interactions do not apply.
The energy difference between the metastable and equilibrium states 
decreases with the density, and is minimum in the saturated regime.
The attraction, hence, is the more effective the highest the density.
We expect the metastable state to have the usual features induced by
diquark condensation: Meissner effect, a gap in the excitation 
spectrum, and color superconductivity.
The crucial question that should be addressed is whether this 
metastable state becomes stable at sufficiently large density at finite 
coupling.

A partial answer is the following: 
assuming that the strong coupling expansion gives meaningful results,
diquark condensation cannot take place at sufficiently large couplings,
whatever the chemical potential. The reason is that, as we have seen,
the energy difference between the state with diquark condensation and 
the stable state remains positive for any value of the chemical potential 
in the strong coupling limit. A correction to the effective potential
of order $1/g$ cannot remove such energy difference if $g$ is 
sufficiently large. Hence, diquark condensation cannot take place
in the strong coupling region. If, on the other hand, diquarks condense
at some finite chemical potential in the weak coupling regime, there must 
be a critical value of the coupling, $g_c$, such that diquark condensation
takes place in some interval of the baryon density for $g<g_c$, but diquarks 
do not condense at any density for $g>g_c$. This means that the interval 
of baryon densities (in lattice units) at which diquark condensation occurs 
as a stable state shrinks as $g$ increases, and vanishes at $g_c$. 
For $g>g_c$ the state
with diquark condensate survives as a metastable state. 

It is interesting to note the different behaviour of the chiral transition:
since in the strong coupling limit a first order phase transition separates
a state with vanishing density and chiral symmetry spontaneously broken 
from a state saturated of quarks, where chiral symmetry is unbroken,
the chiral transition at high density may persist for any value of the
coupling, from the weak to the strong coupling regimes. The peculiarity
of the strong coupling limit is that the chiral transition occurs at
the onset chemical potential
(the value of $\mu$ at which the baryon density starts to be non-zero).

For two colors QCD a stable phase characterized by diquark condensation
has been found \cite{aloisio1,ko:to:sin,morrison,zhit,sp:son:st,sp:to:ver}. 
Two colors QCD has two peculiarities that
make its investigation simpler: 1) the diquark operator is a color singlet
and can be used as an order parameter in the usual sense, and 2) the
fermion determinant is real and there is no sign problem.
The sign problem prevents the applicability of numerical simulations to 
analyze the phase diagram of three colors QCD and, therefore, 
strong coupling techniques are useful to improve our understanding of
real QCD at finite baryon density.

\acknowledgments

V.L. thanks E. Seiler for a useful discussion.
This work has been partially supported by an INFN-CICyT collaboration and
MCYT (Spain), grant FPA2000-1252.
V. L. has been supported by Ministerio de Ciencia y Tecnolog\'{\i}a
(Spain) under the Ram\'on y Cajal program.


\begin{thebibliography}{99}
\bibitem{phen1}
P. Kroll, in the proceedings of the \textit{Workshop on N$^*$ physics and
non-perturbative QCD}, Trento, 1998,
\textit{Few Body Syst. Suppl.} \textbf{11}, 255 (1999). 
\bibitem{phen2}
T. Barillari, proceedings of the \textit{e$^+$ e$^-$ Physics at
Intermediate Energies Workshop}, SLAC, Stanford 2001,
eConf C0110430 (2001) W02.
\bibitem{bailin}
D. Bailin and A. Love, \prep{107}{1984}{325}. 
\bibitem{shuryak}
R. Rapp, T. Sch\"afer, E. Shuryak, and M. Velknovsky,
\prl{81}{1998}{53}.
\bibitem{alford}
M. Alford, K. Rajagopal, and F. Wilczek, \plb{422}{1998}{247}.
\bibitem{son}
D.T. Son, \prd{59}{1999}{094019}. 
\bibitem{schaf:wil}
T. Sch\"afer and F. Wilczek, \prd{60}{1999}{114033}. 
\bibitem{pisarsky}
R.D. Pisarsky and D.H. Rischke, \prd{61}{2000}{051501}; 
\ibid{61}{2000}{074017}. 
\bibitem{cristina}
C. Manuel, \prd{62}{2000}{114008}.
\bibitem{vertex}
W.E. Brown, J.T. Liu, and H. Ren, \prd{61}{2000}{114012};
\ibid{62}{2000}{054016}.
\bibitem{alford:rew} 
M. Alford, \arnps{51}{2001}{131}. 
\bibitem{al:raj:wil}  
M. Alford, K. Rajagopal, and F. Wilczek, \npb{537}{1999}{443}. 
\bibitem{al:ber:raj} 
M. Alford, J. Berges, and K. Rajagopal, \npb{558}{1999}{219}. 
\bibitem{al:bow:raj}
M. Alford, J. Bowers, and K. Rajagopal, \prd{63}{2001}{074016}.
\bibitem{continuity} 
T. Sch\"afer and F. Wilczek, \prl{82}{1999}{3956}.
\bibitem{elitzur}
S. Elitzur, \prd{12}{1975}{3978}.
\bibitem{splittorff}
K. Splittorff, \heplat{0305013}.
\bibitem{wilczek}
F. Wilczek, \npa{663}{2000}{257}.
\bibitem{langfeld}
K. Langfeld, \heplat{0212032}.
\bibitem{aloisio1}
R. Aloisio, V. Azcoiti, G. Di Carlo, A. Galante, A.F. Grillo,
\plb{493}{2000}{189}; \npb{606}{2001}{322}. 
\bibitem{ko:to:sin}
J.B. Kogut, D. Toublan, and D.K. Sinclair, \npb{642}{2002}{181};
\plb{514}{2001}{77}. 
\bibitem{morrison}
J.B.Kogut, D.K. Sinclair, S.J. Hands, and S.E. Morrison,
\prd{64}{2001}{094505}. 
\bibitem{zhit}
J.B. Kogut, A.A. Stephanov, D. Toublan, J.J.M. Verbaarschot, and
A. Zhitnitsky, \npb{582}{2000}{477}.
\bibitem{sp:son:st} 
K. Splittorf, D.T. Son, and M.A. Stephanov, \prd{64}{2001}{016003}. 
\bibitem{sp:to:ver}
K. Splittorf, D. Toublan, and J.J.M. Verbaarschot, \npb{620}{2002}{290}.
\bibitem{svet}
B. Bringoltz and B. Svetitsky, \heplat{0211018}. 
\bibitem{umino}
Y. Umino, \prd{66}{2002}{074501}. 
\bibitem{xiang}
E.B. Gregory, S.-H. Guo, H. Kroger, and X.-Q. Luo, \prd{62}{2000}{054508}.
\bibitem{morel}
H. Kluberg-Stern, A. Morel, and B. Petersson, \npb{215}{1983}{527}.
\bibitem{kawamoto}
P.H. Damgaard, D. Hochberg, and N. Kawamoto, 
\textit{Phys. Lett.} \textbf{158B} (1985) 239.
\bibitem{dagotto}
I. Barbour, N.-E. Behilil, E. Dagotto, F. Karsch, A. Moreo, M. Stone,
and H.W. Wyld, \npb{275}{1986}{296}.
\bibitem{bilic}
N. Bili\'c, F. Karsch, and K. Redlich, \prd{45}{1992}{3228}. 
\bibitem{chavel}
M. Chavel, \prd{56}{1997}{5596}. 
\bibitem{aloisio2}
R. Aloisio, V. Azcoiti, G. Di Carlo, A. Galante, and A.F. Grillo,
\plb{453}{1999}{275}.
\end{thebibliography}
\end{document}